\documentclass[epjCONF]{svjour}
\newcommand{\order}[1]{ \mathcal{O} \left( #1 \right) }
\newcommand{\ave}[1]{\left\langle #1 \right\rangle}
\newcommand{\sqrts}{\sqrt{s}}
  \newcommand{\lqcd}{\Lambda_{QCD}}

\usepackage{graphics}
\usepackage[varg]{txfonts} 
\usepackage[latin1]{inputenc}
\session-title{Hot and Cold Baryonic Matter -- HCBM 2010}
\begin{document}
\title{Scaling of multiplicity and flow: pre-LHC trends and LHC surprises}
\author{Giorgio Torrieri\thanks{\email{torrieri@th.physik.uni-frankfurt.de}}}
\institute{$^a$FIAS,
  J.W. Goethe Universit\"at, Frankfurt A.M., Germany }
\abstract{
We examine the scaling trends in particle multiplicity and flow observables between SPS, RHIC and LHC, and discuss their compatibility with popular theoretical models.  We examine the way scaling trends between SPS and RHIC are broken at LHC energies, and suggest experimental measurements which can further clarify the situation
} 
\maketitle
\section{Before the LHC: logarithms and triangles}
The azimuthal anisotropy of mean particle momentum ( parametrized by it's
second Fourier component $v_2$), thought of as originating from the
azimuthal anisotropy in collective flow (``elliptic flow''), has long been regarded as an
important observable in heavy ion collisions.
The main reasons for this is that elliptic flow has long been thought
to be ``self-quenching'' \cite{v2orig,v2orig2}: The azimuthal pressure gradient
extinguishes itself soon after the start of the hydrodynamic evolution, so the final
$v_2$ is insensitive to later stages of the fireball evolution and
therefore allows us to probe the hottest, best thermalized, and
possibly deconfined phase.

In addition, as has been shown in, elliptic flow
is highly sensitive to viscosity.  The presence of even a small but
non-negligible viscosity, therefore, can in principle be detected by a
careful analysis of $v_2$ data.

Indeed, one of the most widely cited (in both the academic and popular
press) news coming out of the heavy ion community concerns the discovery, at the relativistic heavy ion collider ( RHIC ), of a ``perfect fluid'',
also sometimes referred to as ``sQGP'' (strongly coupled Quark Gluon Plasma)
\cite{whitebrahms,whitephobos,whitestar,whitephenix,sqgpmiklos}.
The evidence for this claim comes from the successful
modeling of RHIC $v_2$ by boost-invariant hydrodynamics.

Going further in our understanding is hampered by the large number of ``free'' (or, to be more exact, poorly understood from first principles) parameters within the hydrodynamic model: While the equation of state is thought to be understood from lattice simulations \cite{jansbook}, the behavior of shear and bulk viscosity is quantitatively not known around $T_c$, where it is expected the temperature dependence could be non-trivial \cite{denicol,bozvisc,bulkvisc1,bulkvisc2}.   The same goes for the large number of second-order transport coefficients.   While we have some understanding of the initial transverse density of the system (its dependence on the transverse coordinate is thought to be either a ``Glauber'' superposition of p-p collisions \cite{glauber,denterria}  or a partonic semi-classical ``color glass'' \cite{cgc}), we do not as yet have control over the degree of transparency of the system, the amount of transverse flow created before thermalization (thought to be necessary to make the data agree with particle interferometry measurements \cite{pratt}), or of the interplay between the ``medium'' and the surrounding ``corona'' of peripheral p-p collisions \cite{corecorona1,corecorona2}.   A model incorporating ``all physics'', therefore, is expected to have a lot of correlated parameters which will be highly non-trivial to disentangle.

A tool with the potential of overcoming these difficulties is scaling naturalness.  
Experiments have collected an extraordinary amount of flow data, encompassing a wide range of Energy ($\sqrt{s}$),centrality (parametrized by number of participants $N_{part}$), system size (mass number $A$ of the nuclei), rapidity $y$\footnote{In this article we interchangeably use the rapidity $y=\tanh(p_z/E)$ and the pseudorapidity $\eta = \tanh(p_z/p)$. Away from mid-rapidity and low $p_T$ the two are the same to a very good approximation. Other than the label of Fig. 1, taken from \cite{busza}, $\eta$ in the text of this paper  refers to viscosity, not to pseudorapidity
}, particle species and transverse momentum ($p_T$).   

The experimental data collected allows us to ``scan'' observables dependence on variables relevant to the theory, and to see if the observable change when the same variable is obtained in different ways (for example, flow at mid-rapidity of a lower $\sqrt{s}$ collision compared with flow at the fragmentation region of a higher $\sqrt{s}$ collision,at the same multiplicity density $dN/dy$).

This work qualitatively examines the scaling of multiplicity with $\sqrt{s}$, and the scaling of flow observables with both $\sqrt{s}$ and $dN/dy$, between $\sqrt{s}=19.6$ GeV (SPS and lower RHIC energies) and $\sqrts=2760$ GeV (LHC energies).

This exercise, even when done in a very qualitative level, can yield significant insights into the dynamics since multiplicity and flow observables's scaling with rapidity and $\sqrt{s}$ were remarkably simple \cite{busza,borglhc,mescaling1,mescaling2}.
     The main trends examined in this work are:
\begin{figure*}
\resizebox{1.99\columnwidth}{!}{%
  \includegraphics{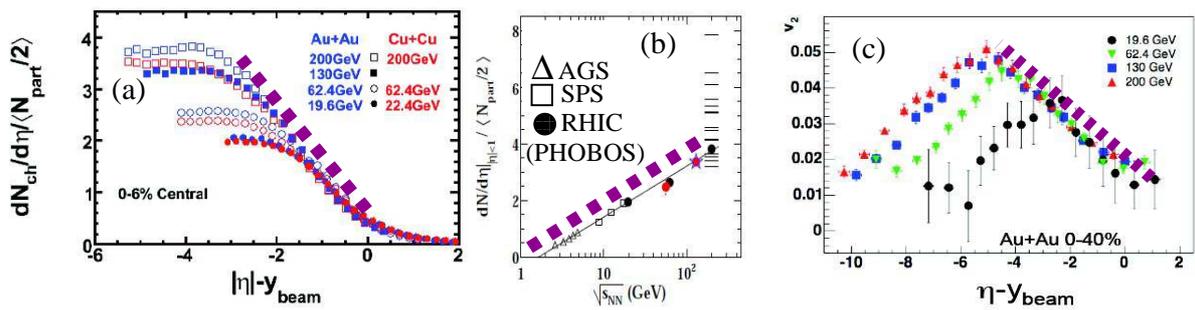} }
\caption{Scalings of multiplicity and flow between SPS and top RHIC energies.  Left panel shows limiting fragmentation, middle panel the logarithmic dependence of multiplicity with energy, and right panel limiting fragmentation of elliptic flow.   Experimental data taken from \cite{busza}
\label{figeta}   }    
\end{figure*}
\label{intro}
\begin{itemize}
\item Limiting fragmentation of $dN/dy$: At high rapidities, the slope of $dN/dy$ becomes independent of reaction energy, the multiplicity curves, plotted w.r.t. $y-y_{beam} = y-\ln\sqrts/2$ can be superimposed 
\item Logarithmic dependence of multiplicity on $\sqrts$: Multiplicity at mid-rapidity scales approximately linearly with number of participants and logarithmically with the center of mass energy $\sqrts$
\item Limiting fragmentation of $v_2$: the limiting fragmentation can also be observed for elliptic flow.  Consequently, central elliptic flow scaled by eccentricity also depends monotonically on $dN/dy$ scaled by the area \cite{mescaling2}
\end{itemize}
Fig. \ref{figeta} summarizes these trends.

The first of these phenomena was noticed a long time ago and is relatively straight-forward to understand \cite{limfrag}:  In the co-moving frame of one of the nuclei, the other nucleus looks like a very thin Lorentz-contracted pancake.
At high enough rapidity, the pancake is thin compared to all other scales of the system (basically the proton size $\lqcd^{-1}$).   Hence, dynamics should not be sensitive to how boosted the frame is w.r.t. pancake as long as the boost is ``large''.  These concepts can be naturally implemented in Regge ``string-based'' models \cite{capella,wernerstring,pajares} and are compatible with the partonic description of hadrons \cite{bgk}.

The question arising naturally is... how large does ``large'' rapidity need to be?   Looking at rapidity distributions, one can see that it can be very small indeed, even at ultra-relativistic energies:  Limiting fragmentation breaks off into a rapidity plateau which, contrary to the predictions of Feynman scaling \cite{feynman,bjorken}, is only $\Delta y \sim 1$ wide, a width that is largely independent of $\sqrt{s}$.
This fact provides a natural explanation for $dN/dy \sim \ln \sqrts$:  Since the total width of the rapidity distribution is $\sim \ln \sqrts$ by kinematics, and the width of the tip of the rapidity distribution is $\sim \sqrts^0$, the height of the rapidity distribution should also  $ \sim \ln \sqrts$.  

This apparently simple ``explanation'' hides a very non-trivial initial state dynamics:  The reasoning used to explain limiting fragmentation, when extended to mid-rapidity, presupposes {\em two} ``pancakes'' passing through each other, {\em each} much thinner than $\lqcd^{-1}$.  Hence, one would expect dynamics to be invariant under boosts, in other words a {\em large} (until $y$ is a large fraction of $y_{lim}$) plateau around mid-rapidity.

This is equivalent to the Feynman description
of a boost-invariant multiplicity distribution \cite{feynman,bjorken} up to ``high'' rapidities $\Delta y \sim \order{y_{lim}} \sim \order{1}\ln\sqrts$, a description natural in an asymptotically free theory, since an interaction moving two high-rapidity partons into mid-rapidity has to be ``hard'' and hence suppressed.   Instead, experimental data says that the rapidity pleateau is either non-existent or small ($\Delta y\sim 1$, of the order of the thermal smearing expectation) and independent of $\sqrt{s}$.  The simplest parton-string models, where string ends predominantly originate from valence quarks, can not explain such a dependence, and hence can not describe $dN/dy \sim\ln \sqrts$

The most natural way of reconciling limiting fragmentation away from mid-rapidity with the absence of boost invariance seems to be \cite{mescaling1} to maintain a ``stringy'' initial state at formation time $\tau_0$, but admit that the spacetime locus of collision (the spacetime region at zero rapidity),rather then the target and the projectile, act as sources of most string ends.   A quantitative model of this type within QCD is, however, to date lacking, through something similar can be achieved by allowing for low $x$ sea partons to form ``small chains'' (separated by very little rapidity) \cite{capella}, or by representing these small $-x$ partons as string excitations \cite{wernerstring}.

If this scenario is correct, then, initially, flow is boost-invariant \cite{feynman,bjorken} (the spacetime rapidity is equal to the flow rapidity), but the initial density strongly depends on rapidity, and hence boost-invariance is {\em not} a good approximation, except very close to mid-rapidity.     A generic prediction of this scenario is the breakdown of limiting fragmentation when $dN/dy$ stops growing logarithmically with  $\ln \sqrts$, something that can be studied at the LHC,as the next section will show \cite{alicemult}.

If, as commonly thought, elliptic flow is hydrodynamic in origin, the density rapidity variation means that at high rapidity most of the system will be hadronic, while at more central rapidity it will be partonic.
Assuming the start of the expansion proper time $\tau_{eq}$ is only weakly dependent on $\sqrt{s}$ (assuming otherwise generally breaks limiting fragmentation \cite{mescaling1}), this seems to happen at $y - y_{beam} \sim 1$ \cite{mescaling1} for asymptotically high energies.  This region has been accessed at both SPS and RHIC energies, and is shown in the distributions of Fig.\ref{figeta}.

Given these considerations, limiting fragmentation of $v_2$ is particularly surprising:
Assuming the rapidity dependence of flow observables is ``encoded'' in the initial density (and associated intensive properties: $T,\eta/s$ etc.) rather than in the transverse size ($S \sim A^{2/3}$ at all $y$), and assuming subsequent evolution is local in $y$, we should expect the scaling of $v_2$ to follow the scaling of $dN/dy$ {\em provided} the intensive properties are either invariant  with the parton density at equilibrium$\rho(\tau_{eq})$ ($\tau_{eq}$ is the time hydrodynamic evolution starts, $\sim$ the mean free path $\l_{mfp}\sim \eta/(Ts)$), or change monotonically with $\rho(\tau_{eq})$, throughout the rapidity range.   However, in a cross-over from a QGP to a hadron gas, intensive properties of the system should {\em not} change monotonically with $\rho(\tau_{eq})$.

\begin{figure*}
\resizebox{1.99\columnwidth}{!}{%
  \includegraphics{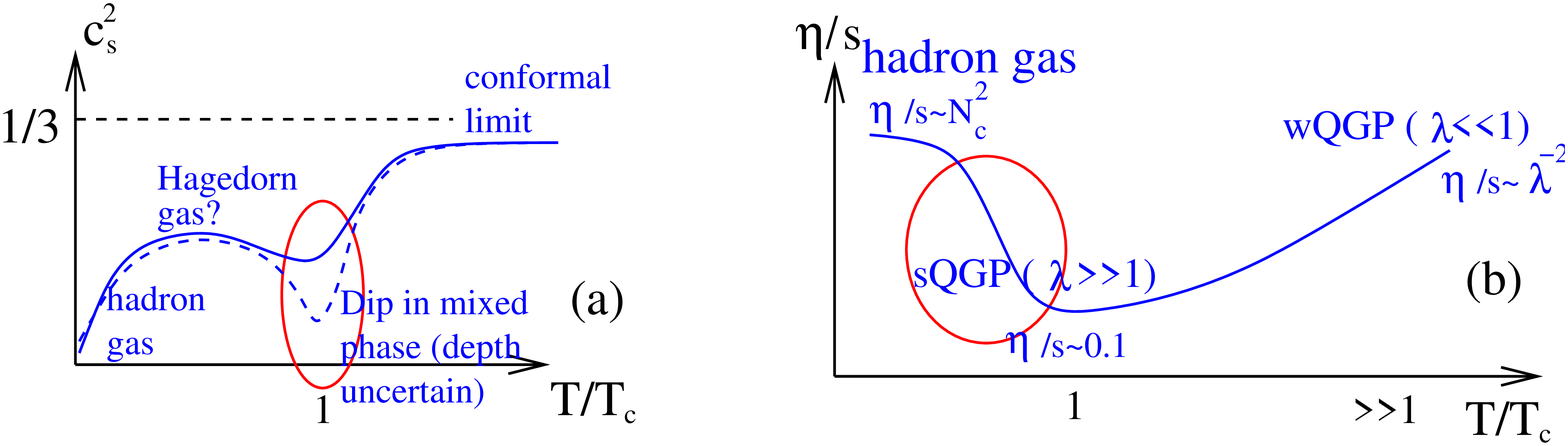} }
\caption{
Breakdown of scaling with temperature of the parameters of the hydrodynamic model
\label{figT}   }    
\end{figure*}
In Fig. \ref{figT}, we summarize the expected changes:  The viscosity to entropy ratio $\eta/s$ is expected to jump from the relatively high value of the hadron gas ($\eta/s \sim N_c^2$ in a gas of mesons and glueballs, where $N_c$ is the number of colors \cite{thooft}) to the low value of strongly interacting QGP ($\eta/s \sim \order{0.1} N_c^0$ \cite{gyulvisc}), and then slowly increase to the asymptotically free weakly coupled QGP ($\eta/s \sim \order{\lambda^{-2}} N_c^0$, where $\lambda$ is the `t Hooft coupling constant\cite{gyulvisc}).  In addition, the speed of sound is expected to have a dip in the cross-over region, whose depth is at the moment not well determined \cite{choj,jansbook}.   

In a wide variety of models, elliptic flow $v_2$ depends on the eccentricity $\epsilon$, and should be approximately proportional to it. This essentially follows from Taylor-expanding the solution of whatever dynamical equation $v_2$ obeys in $\epsilon$, since
 $\epsilon$ is small and dimensionless, and since by symmetry with no $\epsilon$ there is no elliptic flow.   

We also know that $v_2/\epsilon$ decreases if viscosity is turned on, i.e. if $K$ increases.     Hence, it is quite natural that, as suggested in \cite{dumitru,mescaling2}
\begin{equation}
\label{dumeq}
\frac{v_2}{\epsilon}\sim \left. \frac{v_2}{\epsilon} \right|_{ideal} \left(1-\frac{K}{K_0} \right)  \simeq \left. \frac{v_2}{\epsilon} \right|_{ideal} \frac{K^{-1}}{K^{-1}+K_0^{-1}} 
\end{equation}
where $K$ is the Knudsen number and $K_0 \sim \order{1}$ is a parameter specific to the theory.
Furthermore, the transverse Knudsen number at a given mean free path $l_{mfp}$ is
\begin{equation}
\label{khydro}
K^{-1} \sim \l_{mfp}^{-1} \sqrt{S} \sim  \frac{ c_s}{ l_{mfp} S} \frac{dN}{dy}
\label{dumitru}
\end{equation}
This formula assumes just boost-invariant flow, as well as a a time-scale \cite{dumitru} $\tau_{v2}=\sqrt{S}/c_s $ for the building up of 
$v_2$, where $\sqrt{S}$ ($\sim N_{part}^{1/3}$, Not to be confused with the center of mass energy $\sqrt{s}$ or the entropy density $s$) is the initial transverse size of the system (which, as we argued earlier, is independent of rapidity).  
It should be noted that going beyond this rough approximation for $\tau_{v2}$ worsens scaling, since  \cite{v2orig} $\tau_{v2}$ rises {\em and saturates} with increasing density \cite{kestin} due to the self-quenching of elliptic flow.

The derivation of Eqs. \ref{dumeq} and \ref{khydro} follows straight-forwardly \cite{dumitru} from  density formula \cite{bjorken}
\begin{equation}
\label{bjorkdens}
\rho \sim \frac{1}{S \tau_{eq}} \frac{dN}{dy}
\end{equation}
and Taylor expanding

We further remember that $\left. \frac{v_2}{\epsilon} \right|_{ideal}$ depends on the equation of state, i.e. on the speed of sound.
By a leading order expansion argument, and remembering that the asymptotic expansion speed of a Godunov-type hydrodynamic shock wave $\sim c_s$ \cite{landau}, it can be seen that $\left. 
\frac{v_2}{\epsilon} \right|_{ideal} \sim c_s$ (numerical simulations lend credence to this scaling, see \cite{dumitru}).   We also remember that the mean free path $l_{mfp} \sim \frac{\eta}{T s}$. 

Putting everything together, and neglecting the difference between $y$ and the pseudo-rapidity $\eta$ (small in the fragmentation region away from mid-rapidity), we get that
\begin{equation}
\label{v2scalstrong}
\frac{v_2}{\epsilon} \sim c_s(\tau_{eq})\left(1 - \order{N_{part}^{-1/3}\mathrm{fm^{-1}}}\left[  \frac{c_s \eta}{T s} \right]_{\tau_{eq}} \right)
\end{equation}
we believe that when $T>T_c$ $\eta/s \ll 1$, $c_s \simeq 1/\sqrt{3}$, when $T\sim T_c$ $c_s \ll 1/\sqrt{3}$ and $\eta/s$ is at a minimum, and when $T<T_c$ $c_s$ goes back to a value not too different from $1/\sqrt{3}$ but $\eta/s$ increases to $\geq 1$.    (Fig. \ref{figT}).      

Additionally, close to mid-rapidity the plasma lifetime should $\gg \tau_{v2}$, so $v_2$ saturates and becomes independent of initial density (Eq. \ref{v2scalstrong} over-predicts $v_2/\epsilon$).  In the less dense region, however, the plasma lifetime $\leq \tau_{v2}$, so $v_2$ should be approximately proportional to the initial density (Eq. \ref{v2scalstrong} is a good approximation).

On the other hand, $N_{part}$ should be independent of rapidity and pseudorapidity  while the initial $T(\tau_{eq})$ should smoothly change as $\sim \left( dN/dy\right)^{1/3}$ \cite{bjorken}.

\begin{figure}[t]
\resizebox{0.99\columnwidth}{!}{%
  \includegraphics{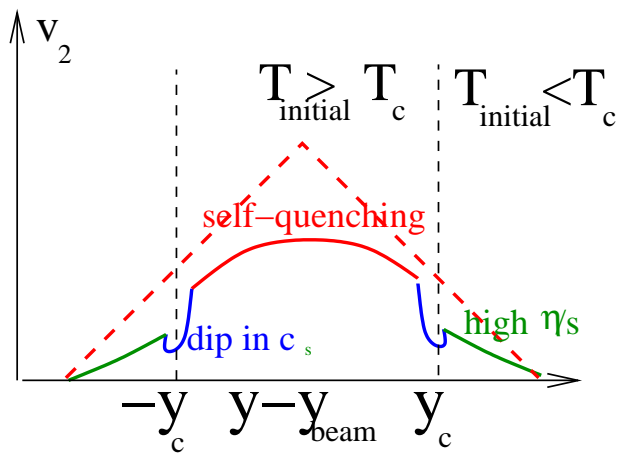} }
\caption{\label{figv2} (color online) The $v_2$ dependence on rapidity given initial conditions reproducing limiting fragmentation, and subsequent 
hydrodynamic evolution. The superimposed dashed line shows the pre-equilibrium ($\tau=\tau_{dyn}$) partonic density motivated by the $dN/dy \sim \ln \sqrts$ and well reproduced by models such as \cite{capella} }
\end{figure}

We immediately see that the scaling seen in the right panel of Fig. \ref{figeta} is {\em not} compatible with a modified BGK initial condition, or indeed any initial condition without an unphysically finely tuned correlation between the size of the system and intensive parameters \cite{mescaling2}.  In the supposedly long-lived ideal fluid mid-rapidity region, $v_2/\epsilon$ 
should be considerably flatter than $dN/dy$ due to the self-quenching of $v_2$.
    At the critical rapidity where $T_{eq} \sim T_c$, $v_2/\epsilon$ should dip due to the dip in the speed of sound, and in the fragmentation regions where $T_0<T_{eq}$ $v_2/\epsilon$ should go down more rapidly than $dN/dy$ due to the rise in $\eta/s$.  The rapidity at which $v_2$ vanishes should in general be different from the rapidity at which $dN/dy$ does, due to $v_2$ additional dependence on $\eta/s$ and system lifetime.  These expected trends are summarized in Fig. \ref{figv2}

These considerations have prompted us \cite{mescaling1} to doubt the hydrodynamic paradigm, and to search for a model where $v_2 \sim dN/dy$ independently of the thermodynamic properties of the system: Note that
at formation time $\tau_0$ ,when dynamics starts, the system is partonic throughout the whole rapidity range  So a far-from equilibrium system expanding very early will be immune from these considerations.  However, hydrodynamics dictates that flow starts at local equilibrium $\tau_{eq}$ much later then the formation time.  Only at $\tau_{eq}$ the system ``knows'' whether it is in the low $\eta/s$ partonic phase or in the hadronic phase.
 Perhaps the most straight-forward way of forming flow before equilibrium is to assume a Knudsen number $\sim 1$, which, as has been shown in \cite{borghini}, can generate a significant amount of $v_2$ (through not enough to describe experimental data):   Since initial conditions are partonic throughout, therefore, $v_2$ can be insensitive to weather the ``equilibrium temperature'' at that rapidity is above or below $T_c$, thus ensuring that $v_2$ depends monotonically on $dN/dy$ \cite{mescaling1,heisel}.   Strong mean fields $\sim dN/dy$ could then conceivably bring $v_2$ up to experimental values without disturbing the scaling \cite{koch}.

The trends described in this section allowed to predict both multiplicity and flow observables at the LHC \cite{busza,borglhc};  It was hoped that this scaling would allow for a rigorous test of hydrodynamics, since the projected increase in flow would place $v_2$ above the hydrodynamic limit.   As usual, however, data confounded facile predictions.

\section{Post LHC: More Multiplicity and same flow  }

Experimental data has a way of spoiling simple and elegant trends!  \cite{alicemult} has shown that multiplicity has grown considerably faster than logarithmically, as $\sim s^{0.11}$, though, it should be noted, far slower than Landau hydrodynamics$\sim s^{1/4}$ \cite{wong} and AdS/CFT shockwave predictions $\sim s^{1/3}$ \cite{yarom}, suggesting that initial transparency is still very high, in line with the stringy models described in the previous section.

   $v_2$, on the other hand, has grown approximately logarithmically \cite{alice};  If one renormalizes the $v_2$ triangle shown in \cite{borglhc} with the ratio of eccentricities in the relevant centrality bins (The eccentricity of 20-30$\%$ centrality event class of Pb-Pb reported in \cite{alice} is $\sim 20-30\%$ higher  than the eccentricity of 0-40$\%$ used in \cite{busza}, as calculated from \cite{hajo} and \cite{denterria}), it is clear $v_2$ is in the region predicted by \cite{busza,borglhc} or slightly below it.   

Thus, the $v_2 (\eta)$ scaling is probably in line with the trends from \cite{busza,borglhc}, {\em but} the multiplicity scaling of \cite{busza} is abundantly broken.  This makes it likely that the integrated $v_2/\epsilon$ vs $(1/S)(dN/dy)$ scaling, holding from AGS to LHC, is also broken.

\begin{figure*}
\resizebox{2.\columnwidth}{!}{%
  \includegraphics{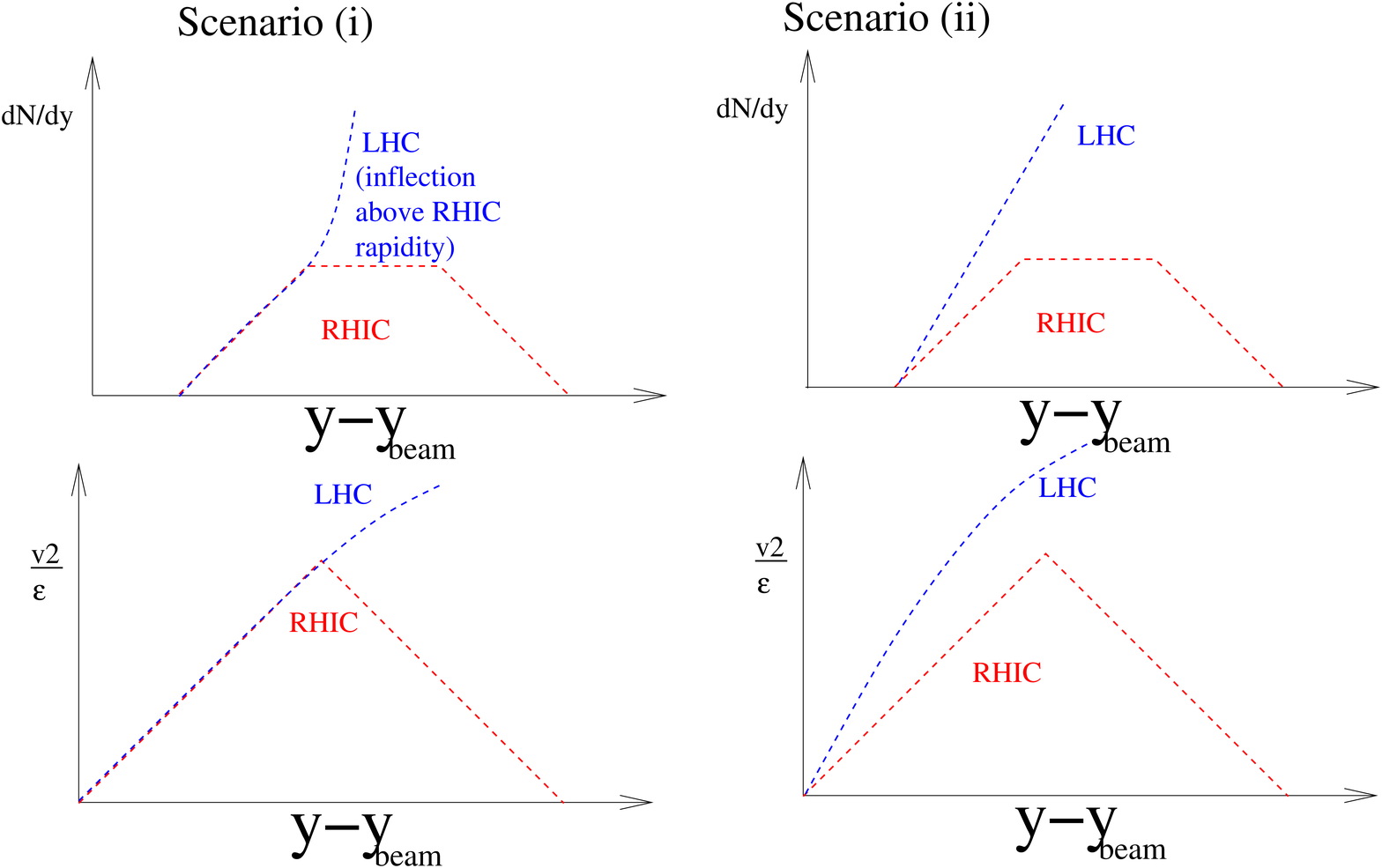} }
\caption{The two interpretions of the trends between RHIC and LHC, and the expected consequences for the rapidity distribution in each scenario.
Scenario (i) presupposes a "soft" background obeying the scaling of \cite{busza}, and an unthermalized extra "jetty" contribution.  Because the jetty contribution is unthermalized, flow maintains limiting fragmentation (bottom panel).  Multiplicity maintains limiting fragmentation up to rapidity smaller than that corresponding to $\sqrt{s}=200$ GeV ($y \simeq 1.3$ ).  In the other limit, limiting fragmentation is broken throughout the fireball's rapidity range due to the contribution of the "extra" multiplicity to flow \cite{mishustiny}.  Reality could be, and probably is, in between these limits 
\label{figfrag}   }    
\end{figure*}

Looking at the preliminary HBT data \cite{lhchbt}, one confirms the $v_2$ results: The nice scaling of $R_{out,side,long}$ with $dN/dy^{1/3}$ \cite{lisa} is most likely broken due to the growth of  $dN/dy$.   While this is not shown explicitly, one can see \cite{lhchbt} that $R_{out}R_{side}R_{long}\sim \ln\sqrts$ (as expected from the old scaling assuming $dN/dy \sim \ln\sqrts$).  {\em however}, $dN/dy$ grows {\em faster} then $\ln\sqrts$.

Its as if ``flow'' observables ($v_2$ and HBT) maintain their old $\ln \sqrts$ scaling, but multiplicity grows faster.
  In this context, one can suggest two ``extreme'' scenarios:
\begin{description}
\item[(i)]  The ``soft'' particle production still follows the scalings suggested in \cite{busza,borglhc}, but the unthermalized minijets contribution to multiplicity can not anymore be neglected when predicting multiplicity.  These minijets, however, are not in thermal equilibrium, and only ``soft'' particle dynamics dictates flow.
\item[(ii)] The ``extra contribution'' above \cite{busza} is fully thermalized and part of the system's soft dynamics.  Flow scales with the total $dN/dy$, but it's scaling is more complex than that suggested by the right panel of Fig. \ref{figeta}.  
\end{description}
These, of course, are extremes, and reality can be somewhere in the middle \footnote{For example, \cite{pajares} seems to find that violations of limiting fragmentation happen at the same time for $dN/dy$ and flow observables}.
A third possibility, that the extra multiplicity reflects the entropy created by the viscous evolution of the not-so-perfect fluid, seems unlikely in light of the good fit obtained of $dN/dy$ as a function of centrality with the most popular initial state models \cite{hijing,alba,alicemult2}; The overwhelming part of the entropy of the system seems to be there from the time of start of the dynamics $\tau_0$.

As fig. \ref{figfrag} shows, a measurement away from mid rapidity, between 
the rapidity corresponding to SPS RHIC energy, (respectively $\sim 2.5$ 
and $\sim 1.3$) at the LHC is essential.   As argued in the previous section, the simplest explanation for the $\ln \sqrt{s}$ scaling of both multiplicity and $v_2$ is that limiting fragmentation holds up to close to mid-rapidity.
If this is true and scenario (i) holds, then, since integrated $v_2$ is dominated by soft particles, limiting fragmentation of $v_2$ should continue (not impossible, given the approximately logarithmic dependence of $v_2$ in \cite{alice}), and limiting fragmentation of $dN/dy$ should gradually break {\em above} the rapidity corresponding to RHIC energy.
On the other hand, scenario (ii) would naturally predict a breaking of limiting fragmentation at all rapidities between RHIC and LHC, since the extra particles produced at mid-rapidity will generate longitudinal and transverse flow \cite{mishustiny}.  Experiment will tell us shortly which is the case.

What do we make of the increase in $v_2$?   Contrary to earlier models, hydrodynamic codes with both viscosity and a long after-burner phase do allow for elliptic flow to rise above the (initially thought to be) ``ideal fluid'' limit of RHIC \cite{divine1}.   The compatibility of such a complex model with simple scalings, however, requires investigation.

To start interpreting these results within the hydrodynamics paradigm, examining a further scaling is necessary: The {\em differential} scaling of $v_2(p_T)$.  As can be seen when comparing \cite{phenix} with \cite{alice,lacey}, below $\sqrt{s} \sim 62$ GeV, $v_2(p_T)$ saturates, and the increase of integrated $v_2$ is driven {\em not} by an increase of $v_2(p_T)$ but by a higher average $\ave{p_T}$.
An ideal hydrodynamics scan \cite{kestin} with different energies produces just such a behavior, {\em provided} $\eta/s$ and $T_{freezeout}$ do not change across the considered $\sqrt{s}$.  A similar simulation with viscous hydrodynamics \cite{songheinz},however, shows that such scaling very quickly breaks down in case $\eta/s$ changes.   

An  interpretation of these findings \cite{lacey} is that $\eta/s$ saturates at $\sqrts=62$ GeV and stays constant up to LHC energies.  But then, what does one make of limiting fragmentation of $v_2$ up to $\sqrts=19.6$ GeV, and of the scaling of  $v_2/\epsilon$ vs $(1/S)(dN/dy)$ scaling between AGS and RHIC energies?

Alternatively, it can be postulated that changes in the initial temperature, $\eta/s$ and $T_{freezeout}$ somehow compensate each other in the final flow observables. Weather this is possible without fine-tuning is still an open question \cite{divine1}.
It should be noted, however, that, while changes in $T_{freezeout}$ are reasonable in hydrodynamics (Since freeze-out happens when the mean free path $l_{mfp} \sim R$, the system size, it is reasonable to assume that the higher the initial energy, the lower $T_{freezeout}$), the data we have seems to disallow such changes over the available $\sqrt{s}$: If this was the case, one would expect a breaking of the scaling of HBT radii with $dN/dy$  (due to longer lifetime in higher density events), as well as a depletion of ratios of particles such as $K^*/K$ with increasing density \cite{usreso1,usreso2}.    It is fair to say no such systematics exists \cite{lisa,na49reso,salur}.

If, as proposed in \cite{lacey}, the $v_2(p_T)$ scaling is the ``fundamental'' one, the critical observable to look for here is $\ave{p_T}(\sqrts)$:  Does $\ave{p_T}$ (thought, unlike $v_2$ to increase uniformly throughout the lifetime of the system) scale, as $v_2$, with $\ln \sqrts$, or does it scale with $dN/dy$ (faster than $\ln \sqrts$ at LHC energies)?  Only in the latter case can we be sure that the increase of total $v_2$ is dominated by $\ave{p_T}$ changes.  The latter case is also favored in scenario (ii) (Scenario (i) would still prefer a parametrically slower growth for $\ave{p_T}$ wrt $dN/dy$ if out-of equilibrium $p_T>2-3$ GeV particles were cut out), and natural in saturation-based scenarios \cite{prasal}.

Independently of these considerations,  the problems related to describing $v_2(y)$ with hydrodynamics will remain.  If, at moderate (corresponding to RHIC $\sqrt{s}$) rapidities limiting fragmentation of $v_2$ is restored, looking for $\ave{p_T}$ and HBT radii dependence of rapidity will clarify to what extent are flow properties correlated with the equation of state.

If flow properties ($v_2,\ave{p_T}$ and HBT radii) continue exhibit limiting fragmentation at moderate rapidity, models of flow generation far from equilibrium such as \cite{borghini,koch} will need to be given serious consideration.

In conclusion, we have reported, and attempted to interpret, experimental scaling trends of multiplicity and flow properties in heavy ion collisions at SPS, RHIC and LHC energies.
Between SPS and RHIC energies both multiplicity and flow observables $\sim \ln \sqrt{s}$ and exhibit limiting fragmentation in rapidity.  While the multiplicity scaling can be naturally described by popular string-based approaches, flow scaling presents a challenge to the generally accepted hydrodynamic model.

Intriguingly, at the LHC, while flow observables seem to scale with $\ln \sqrts$, multiplicity grows faster.  This could mean that a ``hard'' (possibly-nonflowing) component to the ``soft'' (and flowing) background becomes non-negligible.  Determining whether this ``hard'' component flows or not is therefore important.
Examining limiting fragmentation of $v_2$ , $\ave{p_T}$ and HBT radii, and seeing how $\ave{p_T}$ grows with $\sqrts$, can shed light in this direction.
 
I would like to thank Roy Lacey,Pasi Huovinen  and Miklos Gyulassy for discussions and suggestions.
G.T. acknowledges the financial support received from the Helmholtz 
International Center for FAIR within the framework of the LOEWE program
(Landesoffensive zur Entwicklung Wissenschaftlich \"Okonomischer
Exzellenz) launched by the State of Hesse.
We would like to sincerely thank Tamas Biro, Marcus Bleicher and Carsten Greiner for the workshop invitation, and the local organizing committee for their hospitality during the discussions that led to this work.
We would like to thank Columbia University for its hospitality when this work was written up.

\end{document}